\documentclass[fleqn,twoside]{article}
\usepackage[headings]{espcrc2}
\usepackage{graphicx}
\usepackage[figuresright]{rotating}

\newcommand{\mmu}{\mbox{\ensuremath{m_{\mu}}}}

\newcommand{\amu}[1][]{\ensuremath{a_{\mu^{#1}}}}
\newcommand{\gm}{\ensuremath{(g-2)}}
\newcommand{\wa}{\mbox{\ensuremath{\omega_a}}}
\renewcommand{\wp}{\mbox{\ensuremath{\omega_p}}}

\title{Why do we need the new BNL muon $g-2$ experiment now?}

\author{David W. Hertzog\address{Department of Physics \\
University of Illinois at Urbana-Champaign, Urbana, IL 61801
USA}\thanks{Representing the E821~\cite{E821} and the
E969~\cite{E969} Collaborations.}}

\begin{document}
\begin{abstract}
New final results from the CMD-2 and SND $e^+e^-$ annihilation
experiments, together with radiative return measurements from BaBar,
lead to recent improvements in the standard model prediction for the
muon anomaly. The uncertainty at 0.48~ppm---a largely data-driven
result---is now slightly below the experimental uncertainty of
0.54~ppm. The difference, \amu(expt)- \amu(SM) =
$(27.6\pm8.4)\times10^{-10}$, represents a 3.3 standard deviation
effect.  At this level, it is one of the most compelling indicators
of physics beyond the standard model and, at the very least, a major
constraint for speculative new theories such as SUSY or extra
dimensions.  Others at this Workshop detailed further planned
standard model theory improvements to \amu. Here I outline how BNL
E969 will achieve a factor of 2 or more reduction in the
experimental uncertainty.  The new experiment is based on a proven
technique and track record. I argue that this work must be started
now to have maximal impact on the interpretation of the new physics
anticipated to be unearthed at the LHC.

\end{abstract}

\maketitle

\section{Introduction}
At the Tau-04 Workshop, I summarized~\cite{Hertzog-Tau04} the
recently completed Brookhaven E821 muon anomalous magnetic moment
measurements and described a proposal for an extension of the
measurement program.  The E969 experiment~\cite{E969} is designed to
build on the existing infrastructure at BNL by beamline and ring
improvements aimed at increasing the storable muon fraction and,
simultaneously, reducing unwanted hadronic background. The goal is a
reduction in the uncertainty of \amu(expt) by a factor of 2 or more.
Over the past two years,  the proposal plans have been scrutinized
and updated. The technical plan is more complete and a prototype of
one of the new detector concepts has been built and tested. During
the same period, significant progress has been made in the standard
model theory evaluation, as demonstrated here at Tau-06 by many
authors.

The standard model expectation for the muon anomaly is based on QED,
hadronic and weak loop contributions. Passera~\cite{Passera-Tau06}
reviewed the present status and introduced special topics, which
were expanded on by others. Nio~\cite{Nio-Tau06} described an
ambitious effort to complete QED calculations through 10th order.
Vainshtein~\cite{Vainshtein-Tau06} focussed on the evaluation of the
hadronic light-by-light (HLbL) contribution, whose present
uncertainty at 0.2-0.3~ppm will become significant on the scale of
the experimental goals. But the present dominant theory uncertainty
is associated with the data-driven determination of the 1st-order
hadronic vacuum polarization (HVP). In particular, the nagging
question of why the tau hadronic decay data does not agree with the
direct $e^+e^-$ annihilation approach was raised. The problem is
evident in the ``single number'' test to determine the
$\tau^-\rightarrow\nu_{\tau}\pi^{-}\pi^{0}$ branching ratio.  The
tau result is 4.5 standard deviations higher than that predicted by
CVC from $e^+e^-$ data~\cite{Davier-Tau06}. Isospin
corrections~\cite{Lopez-Tau06} and the general QCD compatibility of
tau and $e^+e^-$ data at higher energies~\cite{Maltman-Tau06} were
presented, but no resolution of the problem was offered.

The $e^+e^-$-based HVP situation featured new
results~\cite{Eidelman-Tau06} from the Novosibirsk VEPP-2M
experiments CMD-2 and SND. Both independently measure the $e^+e^-
\rightarrow hadrons$ cross section below the $\phi$ resonance. After
adjusting for several subtle mistakes in the radiative correction
procedures, both experiments are in excellent agreement.
Complementing this direct data is the radiative return method where
the $e^+$ and $ e^-$ beams in a collider are fixed in energy and the
center-of-mass collision energy is reduced by the initial-state
radiation.  To establish the predictive spectrum of such collisions
requires a radiative correction analysis, which is made possible by
the {\tt PHOKHARA} Monte Carlo generator~\cite{Rodrigo-Tau06}.  Data
sets requiring this treatment come from BaBar~\cite{Wang-Tau06} and
KLOE~\cite{Venanzoni-Tau06}. So far, the BaBar results have mainly
focussed on the high-mass region where they have improved knowledge
of various exclusive many-body channels.  The important $\pi\pi$
cross section near the $\rho$ resonance is a particular challenge
since the required precision to be competitive is at the sub-percent
level. KLOE is focussing on just this region and it is well known
that their initial report~\cite{KLOE}, which agrees in the integral
with the direct annihilation measurements, is incompatible in
spectral shape. Venanzone described much larger data sets that have
more robust built-in cross checks (e.g., small- and large-angle
sensitivity to the radiated gamma so the events can be sorted in
different kinematical regions). These data are being
analyzed~\cite{Venanzoni-Tau06} at present and we can expect results
by the next Tau Workshop.

Because the HVP contributions drive the uncertainty on the standard
model prediction, \amu(SM), Davier summarized~\cite{Davier-Tau06}
the situation regarding the hadronic tau decays, radiative return
techniques and the new direct $e^+e^-$ measurements.  His
conclusions on HVP and the other standard model contributions are
compiled in Table~\ref{tab:theory}. The total theoretical
uncertainty is 0.48~ppm, to be compared with 0.54~ppm from the
experiment.  The theory uncertainty is the smaller---and, the work
described here promises that it will be reduced even further in the
coming years.

The comparison of experiment to theory yields
$$ a_{\mu}({\rm expt})- a_{\mu}({\rm SM}) = (27.6 \pm 8.4) \times 10^{-10},$$
a $3.3~\sigma$ effect, which hints strongly of some kind of new
physics contribution. Is there a more significant indicator
anywhere? If we allow that massive neutrinos are now built into an
updated standard model, then the answer is ``no.'' But, this
conclusion carries a number of assumptions. It is based on a
relatively new evaluation of the HVP, which omits the first results
from KLOE using the radiative return method and all the hadronic tau
data~\cite{dehz2}.  The selection made is meant to choose precise,
consistent data sets and it favors the direct annhilation over an
indirect technique that requires corrections.

Why the hadronic tau decay approach disagrees so sharply with the
direct $e^+e^-$ method is not understood, and the subject lingers as
one of the most important unsolved problems that this community
should try to address. The tau-based HVP data determine an \amu~
that is much closer to experiment, so it is clearly important to
understand. We are lead to dismissing it for now by the tau
proponents~\cite{dehz2} who argue that uncertainties in the
treatment of isospin corrections, perhaps manifest by the spectral
shape inconsistency between tau and $e^+e^-$ data, are uncertain at
the required level of accuracy.  Further, the CVC test fails
miserably. Something is not right and the confusion clouds the
stated theoretical \amu\ as quoted.  If the standard model is
accurately represented by the $e^+e^-$ data, then the \gm\ test is
very suggestive of new physics. Based on this interpretation---it is
not a guarantee, only a very good bet---the Collaboration feels
strongly that a compelling motivation to improve the \amu\
measurement exists.  Of course it is imperative to resolve the tau
issue and to corroborate the direct $e^+e^-$ data with at least one
of the radiative return methods from KLOE, BaBar or Belle; but, this
work continues and we can expect results in the near term.

\begin{table*}[hbt]
\caption{2006 Standard model theory and experiment summarized at
this Workshop~\cite{Davier-Tau06}\label{tab:theory}}
\begin{tabular}{lrrcl}
\hline
Contribution & Value  & Error & Reference & Comment \\
             & $\times 10^{10}$ & $\times 10^{10}$ & & \\
\hline
QED & 11\,658\,471.9 & 0.1 & \cite{kinqed,Nio-Tau06} & 4 loops; 5th estimated; see Nio \\
Hadronic vacuum polarization~~~~~~ & 690.9 & 4.4 & \cite{Davier-Tau06} & Only CMD-2 and SND used \\
Hadronic light by light &   12.0 & 3.5 & \cite{hlbl} & Value from Ref.~\cite{davmar} \\
Hadronic, other 2nd order & -9.8 & 0.1 & \cite{dehz2} & \\
Weak & 15.4 & 0.22 & \cite{davmar} & 2 loops \\
\hline Total theory & 11\,659\,180.4 & 5.6 & \cite{Davier-Tau06} & 0.48 ppm \\
Experiment & 11\,659\,208.0 & 6.3 & \cite{Bennett:2006} & 0.54 ppm \\
Expt. - Thy. & 27.6 & 8.4 & -- & 3.3 standard deviations\\ \hline
\end{tabular}
\end{table*}

\section{E821 Summary}

The E821 data taking was completed in 2001 and the final analyses
are all
published~\cite{Carey:1999dd,Brown:2000sj,Brown:2001mg,Bennett:2002jb,Bennett:2004xx}.
A comprehensive summary of the experiment, containing many of the
details of the methods used for data taking and analysis, was
published~\cite{Bennett:2006} and a more general review is also
available~\cite{HertzogMorse2004}. With nearly equally precise \amu\
determinations from positive and negative muon samples, the
CPT-combined final result is $\amu ^\pm = 11\,659\,208(6) \times
10^{-10}$. The statistical and systematic uncertainties are combined
in quadrature.


A precision measurement depends on control of the systematic
uncertainties.  These errors were reduced in magnitude in each of
the running years.  We group the uncertainties into those related to
\wa, the precession frequency extraction, and those related to \wp,
the determination of the average magnetic field. The achieved
uncertainties were 0.21~ppm and 0.17~ppm, respectively (with
numerous sub-categories contributing to each class of error). The
goal for the new experiment is to limit the precession and field
systematic uncertainties each to 0.1~ppm. This will require new data
taking operation modes of the storage ring to reduce some of the
more important beam-dynamics related uncertainties and the
installation of higher-granularity detectors to reduce the pile-up
(instantaneous) rate correction.  The field uncertainty improvements
will be achieved by a continued refinement of the present operation
of shimming, measuring, and monitoring.


\section{How is g-2 measured?}

The modern method to determine the muon anomalous moment is to
measure the difference frequency \wa\ between the spin precession
and the cyclotron motion of an ensemble of polarized muons, which
circulate a storage ring in a highly uniform magnetic field. Apart
from very small corrections, \wa\ is proportional to \amu. Vertical
containment is achieved by electric quadrupoles. Therefore, the muon
spin can be affected by the motional magnetic field. Formally, in
the presence of a magnetic and electric field having $\vec {B} \cdot
\vec {\beta} = \vec {E} \cdot \vec {\beta} = 0$, \wa\ is described
by
\begin{eqnarray}
\label{eq:spin-precess}
 \vec{\omega _{a}}  & \equiv & \vec{\omega}_s - \vec{\omega}_c
  \\
& = & \frac{e}{\mmu c} \left[\amu\vec{B} - \left(\amu -\frac
 {1}{\gamma^2-1}\right)(\vec{\beta} \times \vec{E})\right]\nonumber
\end{eqnarray}
where $\vec{\beta}$ represents the muon direction and $\omega_s$ and
$\omega_c$ are the spin and cyclotron frequencies, respectively. The
term in parentheses multiplying $\vec{\beta} \times \vec{E}$
vanishes at $\gamma = 29.3$ and the electrostatic focussing does not
affect the spin (except for a correction necessary to account for
the finite momentum range $\Delta P/P \approx \pm0.14\%$ around the
magic momentum). Equation~\ref{eq:spin-precess} can be rearranged to
isolate \amu, giving $\wa /B$ multiplied by physical quantities that
are known to high precision.

The magic momentum sets the scale of the experiment and the BNL
storage ring~\cite{ring} is 7.1~m in radius and and has a 1.45~T
magnetic field. At 3.094~GeV/$c$ the time-dilated muon lifetime is
64.4~$\mu$s, and the decay electrons\footnote{By convention, we
discuss negative muons and their decay electrons throughout this
paper.} have an upper lab-frame energy of approximately 3.1~GeV. A
key feature of the experiment is the injection of a short pulse of
polarized muons, which enter the ring tangent to, but offset from,
the central storage radius.  They are kicked transversely to align
with a trajectory concentric with the ring center.

\begin{figure}
\includegraphics*[width=0.95\columnwidth]{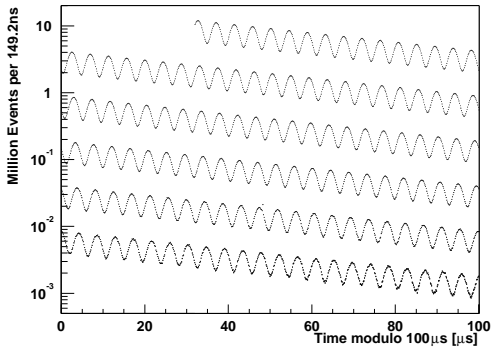} \caption{Distribution
of counts versus time for the 3.6 billion decays in the 2001
negative muon data-taking period. \label{fig:wiggles}}
\end{figure}

The rate of detected electrons having an energy greater than a set
threshold is an exponential (as above), but modulated at the
anomalous precession frequency, \wa, see Fig.~\ref{fig:wiggles}. The
key to the experiment is to determine this frequency to high
precision and to measure the average magnetic field to equal or
better precision.  The field is determined from a suite of NMR
measurements~\cite{nmr,liu} using calibration, fixed and movable
probes to reference the field against a known standard, to monitor
the field continuously, and to map the field in the storage ring
aperture, respectively.  The techniques are documented in
Ref.~\cite{Bennett:2006}.

\section{Steps toward an improved g-2 measurement}
The key to improving on the E821 experiment is obtaining more stored
muons, lowering the hadronic-induced background, and managing the
high instantaneous rate at the time of the fit start, approximately
25~$\mu$s after injection. The plan assumes running at the average
proton intensity of 50~Tp$/$cycle achieved in E821. Although it
introduces no new extraordinary demands on the AGS complex, five
years have passed since high-intensity proton running has occurred
and this mode of operation will have to be re-established. Outlined
below is a relatively conservative set of upgrades for the new
experiment, the core of which forms the theme in the
proposal~\cite{E969}. The more creative ``backward'' decay beam
concept (see also Ref.~\cite{Hertzog-Tau04}) is not described here
because technical and cost uncertainties must be better understood
before we commit to its implementation. Instead, a simple and
largely tested ``forward'' decay scheme, along with several new
ideas that have recognized advantages, are presented.

\subsection{Greater bunch rate}
The ideal experiment divides the accelerator proton budget into a
large number of short-bunch-length pulses per cycle.  Each bunch
generates an independent muon ``fill.''  Twelve bunches per 2.7~s
cycle is the normal operation at the AGS.  With a muon fill lasting
no more than 1~ms, the experiment is ``on'' only 0.44 percent of the
time. The instantaneous rate exceeds several MHz per detector and
already implies a significant pileup correction of a few percent. As
the stored muon rate must increase to achieve the statistical goal,
the pileup fraction must not increase, which can be realized by
subdividing the detectors and/or decreasing the average rate per
bunch.  The latter implies a higher bunch rate per cycle, which
places new demands on the AGS operations. A scheme worked out many
years ago splits the primary 12 bunches spatially into three
distinct co-moving components. Each ``sub-bunch'' is extracted
independently. This operation is not yet a tested concept, and it
likely requires new extraction hardware. But, it is an attractive
idea that would improve the experiment tremendously if it can be
realized.

\subsection{Improved muon collection}
\begin{figure}
  \begin{center}
    \includegraphics*[width=\columnwidth]{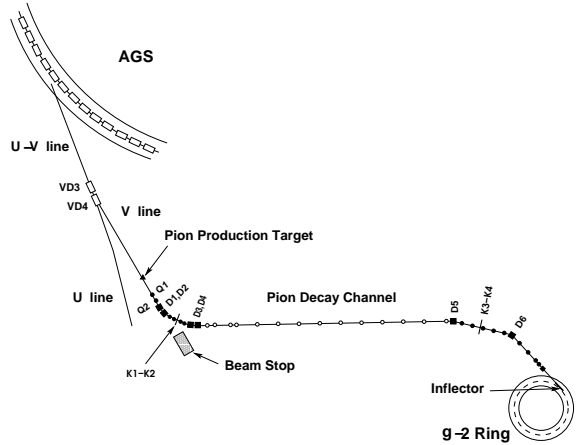}
    \caption{Plan view of the pion/muon beamline. \label{fig:beamline}}
  \end{center}
\end{figure}

The pion-muon beamline upstream of the ring has three components: a
pion creation and collection section; a pion-to-muon decay section;
and a final muon selection at $P_{magic} = 3.094~$GeV/$c$.
Figure~\ref{fig:beamline} is a schematic of the beamline and storage
ring. A number of competing factors enter into final figure-of-merit
(loosely defined as muons into the ring times average polarizition
squared times fraction stored). The numbers given below have all
these factors folded together into a simple ``stored muon'' factor.

The pion collection and muon selection sections are tuned to
slightly different momenta. Typical operating conditions in E821 set
the pion momentum $1.7\%$ greater than $P_{magic}$. Forward decay
muons, but slightly off axis from zero degrees, are contained in the
decay channel and transported through the final bending arc
indicated by dipoles D5 and D6, which are tuned to the $P_{magic}$
momentum. These muons have a longitudinal polarization of $\approx
94.5\%$. The slightly higher momentum pions that enter the final arc
are largely cut at the K3/K4 slit; however, the low-momentum tail of
the pion distribution is transported through the slits and into the
ring. The pion-to-muon ratio is typically 1:1 at nominal settings,
but this ratio can be controlled by many factors.  At the first pion
collection arc, slits K1/K2 define a pion momentum acceptance.  In
operation, a narrow $\Delta P/P$ of $\approx 0.5\%$ was used. Higher
flux can be obtained, but only at the expense of higher pion
contamination at the final K3/K4 selection as well. In contrast,
increasing the $P_{\pi}/P_{magic}$ ratio reduces both the muon and
pion flux entering the ring, with the pion contamination falling
typically faster and therefore improving the muon purity.

Under the best operating conditions, the pion flux still created a
prompt ``hadronic flash'' in E821, which was followed by a slower
neutron capture time lasting 10's of microseconds.  This background
particularly affected the calorimeters placed near the injection
side of the ring and had the effect of limiting the start time of
the physics fits.  The plan described for E969 will decrease the
absolute number of pions entering the ring per fill by increasing
the $P_{\pi}/P_{magic}$ ratio.

\begin{figure}
  \begin{center}
    \includegraphics*[width=\columnwidth]{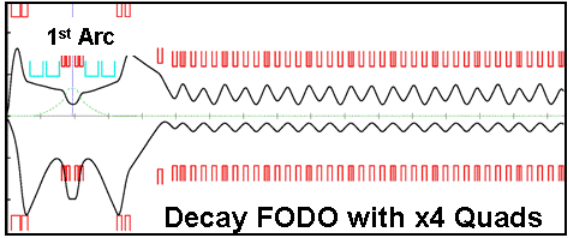}
    \caption{Beamline model of first arc and FODO section with four times
    the present number of quadrupoles.  The envelopes represent the outer contained extent
    of the transported pions.  The top (bottom) panel represents the horizontal (vertical)
    plane. The pions are constrained to lie near the optical axis, which
    leads to a higher fraction of collected decay
    muons.  To fill the phase space of the channel, the accepted muon divergence
    can be larger for a smaller initial spatial displacement.  \label{fig:FODO}}
  \end{center}
\end{figure}

Since the submission of the proposal, we have worked extensively on
improving the beam transport modeling studies, especially in the
80-m long pion-to-muon FODO decay section.  An optimization has been
developed with respect to the hardware changes (more quadrupoles)
and the tuning of momenta and slit acceptances. Our current
conclusions~\cite{Kammel-Pile} suggest that quadrupling the number
of quadrupole elements will help collect more useful muons.
Figure~\ref{fig:FODO} shows a 2nd-order {\tt TRANSPORT} study
indicating the envelope of the transported pions from the 1st arc
through the FODO section. The spatial extent is much smaller
compared to the current line, allowing the collection of a larger
fraction of muons to fill the beamline phase space (essentially, a
larger divergence will be accepted because the spatial deviation of
the decay coordinate is always kept small). Starting with the
assumption that the ratio $P_{\pi}/P_{magic}$ is raised to 1.03, the
effective improvement in muon flux times polarization squared is
approximately 3.3. If estimates of the final ring acceptance are
included, it is reduced to $\approx 2.5$.  But, the pion flux will
have dropped by roughly a factor of 3 for this choice of momentum
ratio. In practice, the final numbers will be tuned to optimize
conditions. The studies leading to these conclusions were drawn from
modeling using {\tt DECAY TURTLE} and {\tt TRANSPORT}, both standard
beam-design tools~\cite{TurtleTransport}.

\subsection{Improved muon transmission through the inflector}
The muon beam enters the storage ring through a superconducting
inflector~\cite{inflector}.  The inflector has coils that
essentially block both the entrance and exit openings.  Studies show
that energy loss and multiple scattering prevents approximately half
of the muons from being stored. Opening the ends of the inflector
will recover these muons. The closed-ended inflector will be
replaced by an open-ended version, which had already been prototyped
and tested prior to the start of E821. The production of a
full-scale open-ended inflector was reviewed during the last two
years and appears to be quite feasible to build.

\subsection{Improved muon storage efficiency}
The muon storage efficiency is largely governed by the kicking and
scraping operations.  In the latter, the quadrupoles~\cite{quads}
are asymmetrically charged to displace the stored beam so that its
outside edges strike round collimators, which define the storage
volume (or aperture).  The scraping process is engaged at the
beginning of a fill and it is turned off in 5-10~$\mu$s.  The more
severe the scraping, the fewer the remaining fraction of muons;
conversely, these muons are well contained, minimizing
time-dependent losses during the fill.  In practice, the scraping
removes roughly 10 percent of the initially stored muons. While
novel ideas have been projected for new scraping schemes, we simply
scale the new experiment with no implied improvement here in rate or
background reduction.

\begin{figure}
    \includegraphics*[width=\columnwidth]{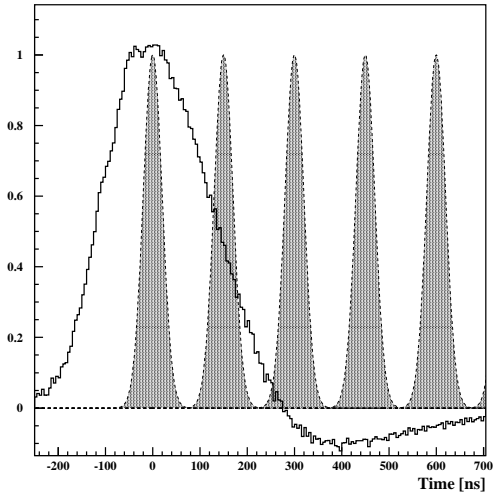}
    \caption{A kicker  pulse. The periodic pulses represent the unmodified muon
    bunch intensity during the first few turns.
    \label{fig:kickwave}}
\end{figure}

An ``ideal'' kicker would provide a perfectly timed $\approx
10$~mrad transverse deflection to the incoming muon bunch one
quarter of a betatron wavelength after the injection. The kicker
perturbation would then vanish in 149~ns, before the muons circulate
the ring once. Improvement in the stored efficiency can be realized
if the actual kicker system~\cite{kicker} can better mimic this
ideal. The E821 device was made of three independent pairs of
parallel current sheets (aluminum plates). Each set is rapidly
energized by an LCR pulse-forming network to induce a high current
and consequent vertical magnetic field. The kicker shape is shown in
Fig.~\ref{fig:kickwave}.  Superimposed is a schematic representation
of the time and width of the muon bunch as it passes the location of
a single kicker section. The kicker shape and magnitude---already a
non-trivial engineering effort to achieve---is evidently too wide.
It is also not strong enough. Consequently, a smaller fraction of
muons is stored and these muons are not perfectly centered by the
kick---a fact that becomes evident in the data where coherent
betatron oscillations appear as an underlying modulation to the
nominal terms required to fit the spectra shown in
Fig.~\ref{fig:wiggles}. Simulations using an ideal kicker pulse
promise a storage efficiency of approximately 8~percent (or more).
The E821 kickers store $\approx 2 - 4$ percent, so there is clear
room for improvement.  This will require revisions of the electrical
charging circuits, especially in the area of reduced inductance (to
narrow the pulse) and some method to increase the maximum plate
voltage.  Clearly, any improvement in the kicker has a direct effect
on the rate and quality of the stored muons. An aggressive R$\&$D
program could likely lead to a factor of 2 in increased stored
muons. The focus would be on the reduction in the inductance at the
pulse-forming network stage (a clear possibility exists) and by
increasing the number of individual kicker sections (either shorter
sections, or one additional full-sized section).

\subsection{Improvements to data taking}
A higher muon rate, and a substantially larger data set, require
rethinking the data-taking strategy.  In E821, 24 lead-scintillating
fiber electromagnetic calorimeters~\cite{detectors} were used to
detect electron energy and time of arrival, each read out by a
single waveform digitizer (WFD). The offline programs found pulses
and corrected for pileup (see Ref.~\cite{Bennett:2006}) A systematic
uncertainty from incomplete pileup subtraction will be exacerbated
at the rates expected for E969; therefore, a new segmented
calorimeter that can isolate simultaneous events and a parallel
method of analysis that is intrinsically immune to pileup, have been
developed. Both will require more modern and higher-rate sampling
waveform digitizers with deep memories.

\begin{figure}
    \includegraphics*[width=\columnwidth]{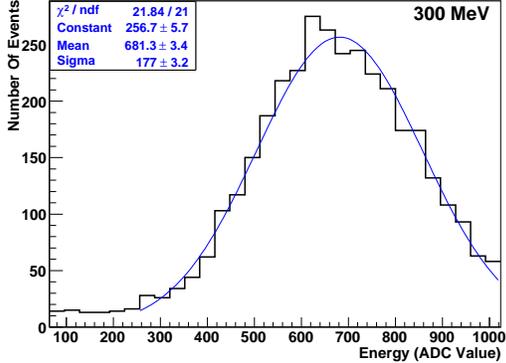}
    \caption{First tests with of a prototype W/SciFi ribbon calorimeter
    using a 300~MeV electron beam at PSI.  The solid line is a Gaussian
    fit to the data.  Its width
    corresponds to a resolution of $\approx 14\%/\sqrt(E)$, which
    was predicted from our {\tt GEANT} simulations.
    \label{fig:calotest}}
\end{figure}

A prototype of the new calorimeter was built and tested. It is made
of alternating layers of 0.5~mm scintillating fiber ribbons and flat
tungsten plates.  The high density yields a Moli$\grave{{\rm e}}$re
radius of $\approx 1.7$~cm and a radiation length of $\approx
0.7~$cm. The compact design is necessary so that the calorimeter
front face can be divided into 20 individual detectors and read out
on the downstream side. {\tt GEANT} studies show that pileup events
can be recognized 4 out of 5 times, which exceeds our requirement.
Results of a brief beam test, using electrons from 125 to 300~MeV at
PSI, show a resolution of approximately $14\%/\sqrt{E}$, in line
with simulations. The detailed balance between improved resolution
(increase the scintillator to tungsten ratio, and/or reduce the
layer thicknesses) and retaining the compact nature of the detector,
continues to be studied. A plot of a 300~MeV electron striking the
detector end-on (i.e, 5-degrees with respect to the fiber axis, as
would be typical in this geometry), is shown in
Fig.~\ref{fig:calotest}.

New waveform digitizers were designed for our PSI ``MuLan``
experiment~\cite{mulan}, which aims to measure the positive muon
lifetime to 1~ppm precision. A system of 340, 500-MHz waveform
digitizers~\cite{bostonwfd} is presently fully operational and
tested. These digitizers can be programmed to meet application
specific goals and we expect to use them as a standard for reading
out the 480 photomultiplier tubes implicit in the E969 plan. In
addition, stable clocks and a complete clock distribution network
are already in hand as is a fully appropriate, high-rate data
acquisition system.

A new method of acquiring data will be carried out in parallel to
our traditional method.  It relies on a simple integration of the
energy flow versus time in each fill. No separate events are
recorded, only the energy absorbed by a calorimeter versus time. The
average electron energy striking a calorimeter is modulated at the
\wa\ frequency because of the association of electron energy with
muon spin direction.  The asymmetry is approximately half that of
the traditional ``event'' method and all events striking the
calorimeter contribute. Detailed {\tt GEANT} simulations indicate
that this data set will be statistically weaker by about 9~percent
compared to the traditional method.  However, it requires no pileup
correction and it is statistically partially independent.  The new
method will provide additional information and a strong cross-check
on the main analysis.

\subsection{Summary of improvements}
The new E969 experiment will require an integrated stored muon flux
roughly 10 times that of E821.  We assume a factor of 5 is required
in the stored muon rate per AGS cycle, coupled with a running period
of approximately 20 weeks.  The improvements outlined increase the
stored muon rate and reduce the hadronic-induced background.  The
instantaneous rate increase must be compensated by a combination of
multi-bunch extraction and detector subdivision.  The summary of
improvements described in terms of expected factors is given in
Table~\ref{tab:improvements}. Notice a stored muon rate increase of
10 is achieved, which exceeds our design goal with a safety margin
of a factor of 2.

\begin{table*}[t]
\caption{Summary of improvements compared to E821, where $N$, $B$
and $R$ represent the overall stored muon figure of merit, the
relative hadron-induced background, and the initial instantaneous
rate, respectively.\label{tab:improvements}}
\begin{tabular}{lcccl}
\hline
Improvement & $N$ & $B$ & $R$ & Comment \\
\hline
AGS bunch increase & 1 & 1 & 0.33 & An attractive possibility, but uncertain\\
Quadruple FODO quads & 2.5 & 0.33 & 2.5 & Detailed studies at $P_{\pi}/P_{magic}=1.03$\\
Open inflector end & 2 & 1 & 2 & Preliminary studies support conclusion\\
Kicker & 2 & 1 & 2 & Simulations in progress; engineering required\\
 \hline
Summary & 10 & 0.33 & 3.3 & \\
\hline
\end{tabular}
\end{table*}

\section{Summary}
Why do we need the new BNL muon g-2 experiment now?  The reasons are
clear:
\begin{itemize}
  \item The experiment can be improved by at least a factor of 2 but it must
be started ``now'' to be ready when the results from the LHC will
demand additional constraints. Several years of R$\&$D and
construction are required before running and analysis can begin.  We
estimate roughly $5 - 6$ years from project start to achieve the
goal.

  \item The standard model theory uncertainty is already slightly smaller
  than experiment and it should be halved again over the next few years.
  The improvements have been driven by the fact that real measurements
  of \amu\ were also being made.  The momentum should be sustained and
  new efforts, especially related to the difficult hadronic
  light-by-light contribution, must be encouraged.  It is quite likely that
  the next sophisticated step will employ lattice QCD techniques.

  \item We are already at a compelling moment.  The present ($e^+e^-$-based
  theory) is 3.3 standard deviations from the experiment, providing
  a strong hint of new physics. If the current discrepancy
  of $27.6 \times 10^{-10}$ persists, with halved experimental and theoretical
  uncertainties, the significance will rise to $6.7~\sigma$.

  \item For specific models, such as SUSY, \amu\ is particulary effective at
  constraining $\tan\beta$---the ratio of Higgs vacuum expectation values---for
  a given superparticle mass. This information will be complementary to the
  anticipated LHC new-particle spectra and it will be crucial in the effort to pin down the
  parameters of the theory behind them.

  \item Independently of SUSY---we do not suggest or depend on this or
  any other specific model as being correct---measuring \amu\ to
  very high precision will register an important constraint for any new
  physics theory to respect.  Some models will predict a large \gm\
  effect, while others will not.  It is information that can likely
  help diagnose new physics.

  \item On the practical side, the project is based on a proven track record by
  an existing team of experts and new enthusiastic collaborators.
  It is efficient to mobilize the Collaboration now while the ring and beamline
  facilities can be dependably re-commissioned, and while the diverse
  expertise exists.
\end{itemize}

\section{Acknowledgments}
The \gm~ experiment is supported in part by the U.S. Department of
Energy, the U.S. National Science Foundation, the German
Bundesminister f\"{u}r Bildung und Forschung, the Russian Ministry
of Science, and the US-Japan Agreement in High Energy Physics.

\end{document}